\documentclass[journal=apchd5,manuscript=article]{achemso}

\usepackage{chemformula} % Formula subscripts using \ch{}
\usepackage[T1]{fontenc} % Use modern font encodings
\usepackage{xcolor}
\usepackage{setspace}
\usepackage{graphicx}
\graphicspath{ {./figures/} }
\usepackage{subcaption}
\usepackage{amsmath}
\usepackage{siunitx}
\usepackage{layouts}
\usepackage{color,soul}
\usepackage[numbers]{natbib}

\author{C\'edric Lemieux-Leduc, Mahmoud R. M. Atalla, Simone Assali, Sebastian Koelling, Patrick Daoust, Lu Luo, G\'erard Daligou, Julien Brodeur, St\'ephane K\'ena-Cohen, Yves-Alain Peter, and Oussama Moutanabbir}
\email{ oussama.moutanabbir@polymtl.ca} %% email address is required; see note below about the corresponding author designation
\affiliation{Department of Engineering Physics, Polytechnique Montr\'eal, C.P. 6079, Succ. Centre-Ville, Montr\'eal, Qu\'ebec, Canada H3C 3A7}

\title{Transfer-printed multiple Ge$_{0.89}$Sn$_{0.11}$ membrane mid-infrared photodetectors}

\begin{document}

\maketitle

%%%%%% Abstract %%%%%%
\begin{abstract}

Due to their narrow band gap and compatibility with silicon processing, germanium-tin (Ge$_{1-x}$Sn$_x$) alloys are a versatile platform for scalable integrated mid-infrared photonics. These semiconductors are typically grown on silicon wafers using Ge as an interlayer. However, the large lattice mismatch in this heteroepitaxy protocol leads to the build-up of compressive strain in the grown layers. This compressive strain limits the material quality and its thermal stability besides expanding the band gap, thereby increasing the Sn content needed to cover a broader range in the mid-infrared. Released Ge$_{1-x}$Sn$_x$ membranes provide an effective way to mitigate these harmful effects of the epitaxial strain and control the band gap energy while enabling the hybrid integration onto different substrates. With this perspective, herein strain-relaxed Ge$_{0.89}$Sn$_{0.11}$ membranes are fabricated and subsequently transfer-printed with metal contacts to create multiple photodetectors in a single transfer step. The resulting photodetectors exhibit an extended photodetection cutoff reaching a wavelength of $3.1 \,\mu$m, coupled with a significant reduction in the dark current of two orders of magnitude as compared to as-grown photoconductive devices. The latter yields a reduced cutoff of $2.8 \,\mu$m due to the inherent compressive strain. Furthermore, the impact of chemical treatment and annealing on the device performance was also investigated showing a further reduction in the dark current. The demonstrated transfer printing, along with the use of an adhesive layer, would allow the transfer of multiple GeSn membranes onto virtually any substrate. This approach paves the way for scalable fabrication of hybrid optoelectronic devices leveraging the tunable band gap of Ge$_{1-x}$Sn$_x$ in the mid-wave infrared range.

\end{abstract}

%%%%%%%%%%%%%%%%%%%%%%%%%%  body  %%%%%%%%%%%%%%%%%%%%%%%%%%

\section*{Introduction}

The mid-wave infrared (MWIR, $2-8 \,\mu$m) range is a highly strategic portion of the electromagnetic spectrum relevant to numerous applications in sensing, imaging, and free-space communication~\cite{lin2018mid, popa2019towards, razeghi2014advances, sieger2016toward}. Current commercial platforms for MWIR detectors predominantly rely on III-V semiconductors such as InSb~\cite{alimi2019insb} and type-II superlattices (T2SL)~\cite{muller2020thermoelectrically, alshahrani2022emerging} and II-VI compounds including HgCdTe (MCT)~\cite{rogalski2020hgcdte, li2020hgcdte}. InSb and MCT detectors cover wavelengths up to $7 \,\mu$m and $12 \,\mu$m, respectively, which enabled the development of MWIR imaging systems based on focal plane arrays (FPAs)~\cite{pusino2016insb, lei2015progress}. While these technologies are mature, they remain costly, suffer from limited integration with silicon-based electronics, and are associated with a low bandwidth that is usually below 10 GHz for photodetectors with a cutoff wavelength above $2.3 \,\mu$m~\cite{chen2021recent}. Recently, germanium-tin (Ge$_{1-x}$Sn$_x$) alloys have emerged as alternate narrow-band gap materials for silicon-compatible MWIR optoelectronics~\cite{moutanabbir2021monolithic}. Indeed, this class of semiconductors covers a broad range of band gap energies by controlling the lattice strain and Sn content~\cite{moutanabbir2021monolithic}. Ge$_{1-x}$Sn$_x$ materials have been used in photodetectors operating at room temperature in the near-infrared (NIR) and MWIR~\cite{tran2019sibased, li202130, talamas2021cmos, atalla2022highbandwidth, liu2022sn}, as well as for modulation based on the Franz-Keldysh effect~\cite{hsieh2021electro}. Additionally, Ge$_{1-x}$Sn$_x$ alloys also pave the way towards silicon-compatible light sources. Indeed, as the atomic content of Sn increases, the band gap at the $\Gamma$-valley diminishes faster than the one at the L-valley, leading to a transition to a direct band gap when the Sn content exceeds 10 at.\% Sn in a relaxed layer. This band gap directness has been exploited to implement silicon-integrated light emitting diodes~\cite{stange2017shortwave, chang2022midinfrared, atalla2023extended} and lasers~\cite{chretien2019gesn, kim2022enhanced, zhou2022electrically, chretien2022room, buca2022roomtemperature}. However, due to the low solubility of Sn in Ge (< 1 at.\%) the growth of high-quality Ge$_{1-x}$Sn$_x$ layers with the minimal Sn composition required for MWIR applications poses challenges. Furthermore, the alloy suffers from a substantial lattice mismatch induced by the 14.7\% difference between Ge and Sn atomic radii, resulting in compressively strained layers. This strain build-up leads to an increase in the band gap energy at the $\Gamma$-valley and impedes the Sn incorporation during growth, thus shrinking the wavelength range covered by the compressively strained devices~\cite{assali2019enhanced}.

Several approaches were investigated to alleviate these drawbacks~\cite{nawwar2022impact}, including the growth of graded Ge$_{1-x}$Sn$_x$ buffers where the composition of Sn is gradually increased by controlling the temperature and precursors flow during growth~\cite{aubin2017growth, dou2018investigation, assali2018atomically}. This method was found to be effective in partially relaxing the compressive strain in the top layer by allowing the misfit dislocations to glide in the initially grown Ge$_{1-x}$Sn$_x$ buffer layers. Ge$_{1-x}$Sn$_x$ nanowires offer an alternative method to relax strain at the facets without compromising the crystalline quality, enabling nanoscale optoelectronic devices with promising performance~\cite{meng2016coreshell, assali2017growth, luo2022extended, assali2019strain, kim2023short, luo2023mid}. For applications requiring larger active material areas, strain-relaxed membranes emerge as a promising solution to mitigate the harmful effects of compressive strain~\cite{rogers2011synthesis, an2023recent}. Material properties that are practically unattainable by conventional growth or wafer bonding techniques can be achieved through the transfer of membranes onto a wide variety of substrates using transfer printing~\cite{yoon2015heterogeneously, linghu2018transfer}. While group-IV~\cite{yuan2006flexible, nam2011strained} and III-V~\cite{liu2017photonic} semiconductor membranes have been introduced for broadband photodetection and light emission, their operational range does not cover the MWIR. Although progress has been made in membranes based on InAs/InAsSb T2SLs~\cite{zamiri2017antimonide} and ultrathin MCT layers~\cite{pan2023vanderwaals}, Ge$_{1-x}$Sn$_x$ alloys provide a solution to implement transferable group-IV MWIR membrane-based photodetectors~\cite{tai2020strain, an2020modulation, atalla2021all, an2021flexible, chen2022transferable}. The fabrication of Ge$_{0.83}$Sn$_{0.17}$ membrane photodetectors resulted in an impressive increase in the detection cutoff from $3.5 \,\mu$m to $4.6 \,\mu$m while maintaining operation at room-temperature~\cite{atalla2021all}. Moreover, Ge$_{1-x}$Sn$_x$ membranes provide a reduced dark current as the conductive Ge and Si layers underneath are removed, which helps in improving the power dissipation and allows the use of a higher bias voltage. However, these membranes suffered from a structural bowing that reduces their adherence to the host substrate after using transfer printing. Alleviating these limitations requires revisiting the growth process to minimize the compressive strain in addition to introducing an adhesive layer~\cite{menard2005bendable} and reducing the number of post-processing steps by adding transferable contacts~\cite{liu2018approaching}.

Herein, to demonstrate the mass-transfer of GeSn membranes, we fabricated Ge$_{0.89}$Sn$_{0.11}$ membrane metal-semiconductor-metal (MSM) photodetectors that operate at room temperature with a cutoff of $3.1 \,\mu$m and a reduced dark current. The growth of a single Ge$_{0.89}$Sn$_{0.11}$ layer led to membranes with a reduced bowing and an initial compressive strain of -0.6\% before the membrane release. The reduced bowing helps in transferring membranes efficiently as an array at a high yield while the relaxation of the residual compressive strain aids extend the photodetection cutoff. To streamline the processing steps, transferable contacts are initially exfoliated and affixed to an elastomeric stamp before the transfer printing step, ensuring their attachment to the membranes. This results in an array of contacted devices upon completion of the transfer process. SU-8, which is used as an adhesive and an insulating layer, is spin-coated on the host substrate before transfer printing. Subsequently, the performance of the membrane devices is systematically analyzed and compared to their as-grown counterparts, followed by the discussion of the influence of chemical treatment and annealing on device characteristics.

\section*{Results and discussion}

\noindent {\bf Growth and characterization of GeSn epilayers} 

The Ge$_{0.89}$Sn$_{0.11}$ alloy growth was carried out in a low-pressure chemical vapor deposition (CVD) reactor on a 4-inch Si (100) wafer. First, a 1.02 $\mu$m-thick Ge virtual substrate (Ge-VS) was grown using $10 \%$ monogermane (GeH$_4$) and a continuous flow of pure hydrogen following a two-temperature growth protocol at \qty{460}{\celsius} and \qty{600}{\celsius}. In addition to reducing the lattice mismatch between the Si substrate and GeSn, Ge-VS also acts as a sacrificial layer, facilitating the GeSn membranes underetching and release.

\begin{figure}[!htb]
    \centering
    \includegraphics[width=\textwidth]{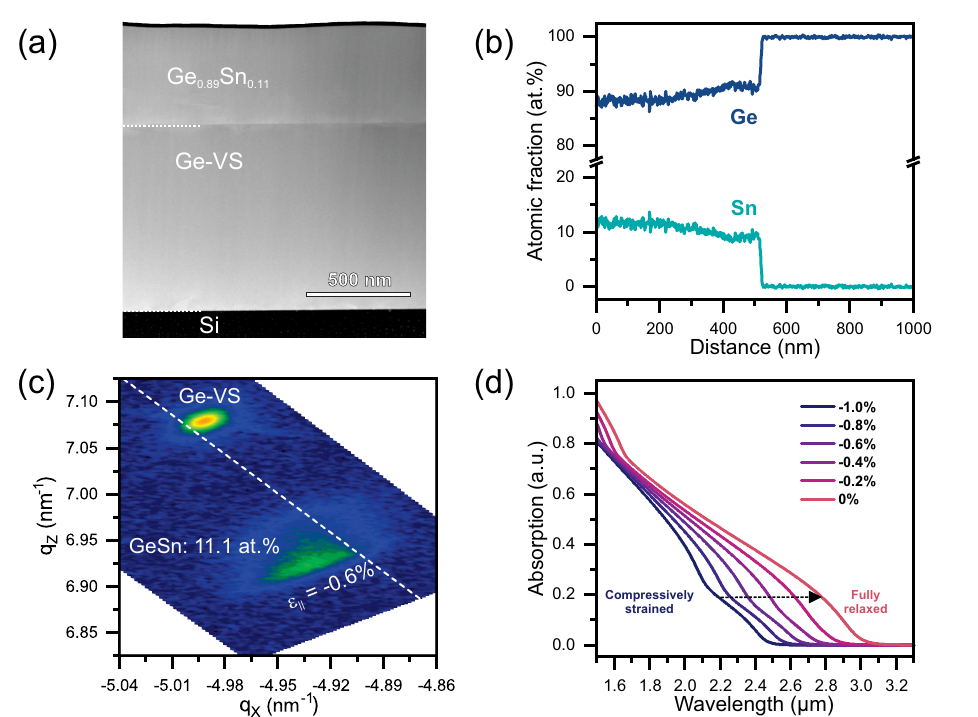}
    \caption{Properties of the as-grown Ge$_{0.89}$Sn$_{0.11}$ alloy. (a) Cross-sectional TEM image of the layer stack. (b) EDX Ge and Sn profiles in the grown layer, starting from the top of the GeSn layer. (c) XRD-RSM around the asymmetrical (224) reflection peak, providing the composition and strain in the as-grown Ge$_{0.89}$Sn$_{0.11}$ layer. (d) 8-band $k\! \cdot\! p$ calculations of the room-temperature absorption of the Ge$_{0.89}$Sn$_{0.11}$ layer by varying the in-plane strain from compressive strain (-1\%) to full relaxation (0\%).}
    \label{fig:1}
\end{figure}

Subsequently, a Ge$_{0.89}$Sn$_{0.11}$ film was grown on Ge-VS at \qty{320}{\celsius} using GeH$_4$ and SnCl$_4$ precursors. The microstructure of the as-grown material was assessed using cross-sectional transmission electron microscopy (TEM), as displayed in Fig. 1(a). Notably, a high density of threading dislocations and defects is localized at the Ge-VS/Si and Ge$_{0.89}$Sn$_{0.11}$/Ge-VS interfaces, indicative of misfit dislocation gliding~\cite{aubin2017growth,assali2019enhanced} while the top of the Ge$_{0.89}$Sn$_{0.11}$ layer is relatively defect-free. To obtain a more detailed understanding of the alloy's elemental composition, energy-dispersive X-ray spectroscopy (EDX) is employed. Fig. 1(b) exhibits the measured Sn and Ge composition profiles, revealing the approximate atomic fraction of Ge and Sn across the Ge$_{1-x}$Sn$_x$ layer. The layer thickness is estimated to be approximately 520 nm and the Sn composition starts at 9\% as it increases towards 12\% from the bottom to the top of the Ge$_{1-x}$Sn$_x$ layer respectively. The background noise coming from the surrounding elements during the EDX measurements was subtracted to avoid an overestimation of the Sn content.

For a comprehensive analysis of the composition and strain in the sample, X-ray diffraction reciprocal space mapping (XRD-RSM) is performed around the asymmetrical (-2-24) reflection peak, as depicted in Fig. 1(c). The map reveals peaks related to the Ge-VS and the GeSn layers. Using the GeSn peak, a composition of ~11.1\% Sn and an in-plane biaxial compressive strain $\varepsilon_{\parallel}$ of -0.6\% are extracted. This composition value is within the range observed in the Sn profile from EDX in Fig. 1(b) and will be used throughout this study.

Fig. 1(d) depicts calculations of the absorption spectrum based on the semi-empirical 8-band $k\! \cdot\! p$ theory~\cite{daligou2023radiative} as the compressive strain is varied from -1\% to 0\% at 300 K. Bowing parameters are implemented in the model to reduce the relative error in the estimation of the band gap at the $\Gamma$ and L symmetry points using Vegard's law. Strain-independent values of 2.85 eV at the $\Gamma$ point and 1.23 eV at the L point were used for the bowing parameters in these calculations. Expectedly, a remarkable increase in the absorption wavelength cutoff is observed as the band energy shrinks due to strain relaxation.

\noindent {\bf GeSn membrane fabrication and transfer printing}

The fabrication of Ge$_{0.89}$Sn$_{0.11}$ membranes starts with a standard photolithography of the as-grown sample to pattern an array of $20 \,\mu$m $\times$ $20 \,\mu$m squares with a pitch of  $200 \,\mu$m $\times$ $400 \,\mu$m, as shown in Fig. 2(a). These patterns are subsequently transferred onto the Ge$_{0.89}$Sn$_{0.11}$ sample by wet etching of GeSn and Ge-VS layers using a mixture of ceric ammonium nitrate and nitric acid. The wet etching process induces an undercut of approximately $1 \,\mu$m with tapered sidewalls, reaching a depth of $1.4 \,\mu$m. By selectively etching the Ge-VS, membranes are formed and fully released from the Si substrate using a CF$_4$-based plasma dry etch. The selectivity in etching Ge-VS rather than Ge$_{1-x}$Sn$_x$ arises from the reaction of fluoride-based radicals with Sn at the exposed surface of Ge$_{1-x}$Sn$_x$. This reaction produces a thin, non-volatile SnF$_y$ layer, effectively shielding the layer from further etching and ensuring a high selectivity~\cite{gupta2013highly}. Images of a single released Ge$_{0.89}$Sn$_{0.11}$ membrane are shown in Fig. 2(b) using optical microscopy (left) and scanning electron microscope (SEM) (right). A slight downward bowing can be observed in the SEM image, indicating either remaining traces of the Ge-VS layer at the center of the membrane or a partial etching of the edge compared to the outer contour of the membrane.

\begin{figure}[!htb]
    \centering
    \includegraphics[width=\textwidth]{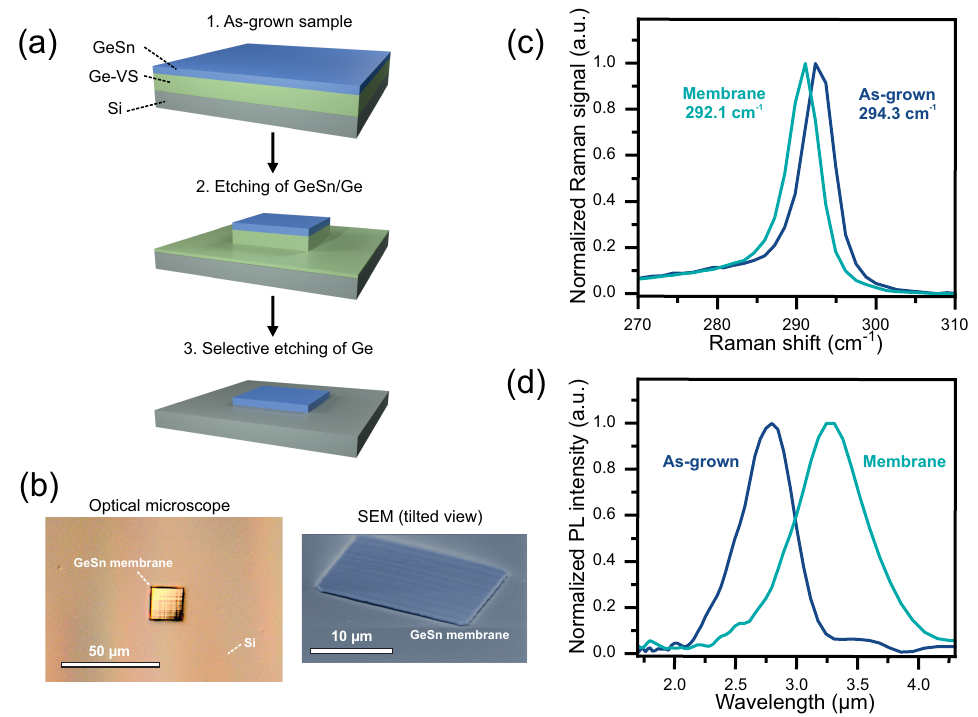}
    \caption{Membrane fabrication. (a) Illustration of the Ge$_{0.89}$Sn$_{0.11}$ membrane fabrication steps. (b) Optical microscope image and SEM micrograph of a released Ge$_{0.89}$Sn$_{0.11}$ membrane. (c) Raman spectroscopy measurements of the as-grown and the membrane samples with the position of the Ge$-$Ge longitudinal phonon Raman peak indicated. (d) Photoluminescence spectra of the as-grown and membrane samples acquired at room temperature.}
    \label{fig:2}
\end{figure}

To assess the impact of the underetching process on strain relaxation, room-temperature Raman spectroscopy measurements were conducted on both as-grown and membrane samples. The acquisition of the Raman spectra shown in Fig. 2(c) was carried out using a 633 nm excitation laser and a $100 \times$ microscope objective lens. The Ge$-$Ge longitudinal (LO) phonon Raman peak is located around $300$ cm$^{-1}$ in pure Ge and it can either red-shift as the Sn concentration increases or blue-shift as the compressive strain increases~\cite{bouthillier2020decoupling}. The peak position shifts from $294.3$ cm$^{-1}$ in the as-grown alloy to $292.1$ cm$^{-1}$ in the membrane, which indicates a relaxation of the compressive strain. Here, a fit with an exponentially modified Gaussian function was performed to extract the central frequency of the asymmetrical Raman peak~\cite{bouthillier2020decoupling}. It is important to note that the released membranes may be subject to local strain fluctuations~\cite{atalla2021all}, which should impact the optical properties as the band structure is locally modified. Room-temperature photoluminescence (PL) analysis was carried out on the investigated samples using an 808 nm excitation laser. The PL peak serves as an approximate indicator of the band gap energy and the absorption range. A $40 \times$ microscope reflective objective lens with a 7 $\mu$m spot size was used to focus the light on a single Ge$_{0.89}$Sn$_{0.11}$ membrane. The spectrum of the as-grown sample depicts a single peak centered at $2.8 \,\mu$m while the membrane shows a single peak at $3.3 \,\mu$m. This redshift of the PL peak indicates a reduction of the band gap energy in the released membrane compared to the as-grown sample, which is in qualitative agreement with the $k \cdot p$ calculations (Fig. 1(d)), showing that the band gap is expected to shrink as the compressive strain is relaxed.

The fabrication of Ge$_{0.89}$Sn$_{0.11}$ membrane photodetectors was carried out using a transfer printing technique~\cite{yoon2015heterogeneously, linghu2018transfer}, as illustrated in Fig. 3(a). This technique utilizes a setup equipped with a microscope and mechanical stages within a glovebox with a nitrogen environment. An elastomeric PDMS stamp, sized roughly \qty{2}{\milli\meter} $\times$ \qty{2}{\milli\meter}, enables the simultaneous pick up of multiple membranes. To create photoconductive detectors, metal contacts are first detached from a separate substrate using mechanical exfoliation. These contacts are then affixed to a PDMS stamp before picking up the membranes. In this approach, pre-processed exfoliated contacts are utilized since the adhesive layer (SU-8) can be easily removed during a metal lift-off process. This technique is versatile and can be employed with various substrates without subjecting them to potentially damaging process conditions. A transferable array of metal contacts with a 5 $\mu$m contact separation gap is patterned in a photoresist layer coated on a separate SiO$_2$/Si substrate using photolithography. Subsequently, a 200 nm-thick Au layer is deposited via electron beam evaporation followed by a lift-off process (Fig. 3(a), step 1.1). The substrate with the Au contacts is then coated with a polycarbonate layer (Fig. 3(a), step 1.2) to facilitate the mechanical exfoliation of the contacts (Fig. 3(a), step 1.3) and enhance their adherence to the PDMS stamp (Fig. 3(a), step 1.4). The donor substrate, which has the released Ge$_{0.89}$Sn$_{0.11}$ membranes, is positioned beneath the PDMS stamp with alignment marks aiding in aligning the membranes between the contact pads (Fig. 3(a), step 2.1). The membranes are then picked up by applying pressure on the sample using the PDMS stamp with the exfoliated contacts (Fig. 3(a), step 2.2). Subsequently, the PDMS stamp is lifted back to its original position to complete the membrane pick up (Fig. 3(a), step 2.3). Afterward, the Si receiver substrate coated with an SU-8 layer is placed beneath the PDMS stamp containing the contacted devices (Fig. 3(a), step 3.1). To transfer the contacted devices, pressure is applied using the PDMS stamp onto the receiver substrate (Fig. 3(a), step 3.2) and the temperature is increased to \qty{200}{\celsius} to melt the polycarbonate film, facilitating the separation of devices from the PDMS stamp (Fig. 3(a), step 3.3). The transfer process is finalized by dissolving the polycarbonate film in dichloromethane for 30 s (Fig. 3(a), step 3.4). To enhance the quality of the Au contacts, the devices undergo annealing at \qty{280}{\celsius} for \qty{30}{\minute}.

Fig. 3(b) presents a microscope image of transferred devices with close-up views from SEM micrographs of representative devices. A noticeable misalignment of contacts with membranes is observed, attributable to both the limited alignment accuracy of the setup and the displacement of the released GeSn membranes from their initial array position during underetch release. The latter can potentially be resolved by an improved design of the membranes or by adding mechanical anchors, which can hold membranes after their complete underetching and can be easily broken during the pick up step~\cite{haq2020microtransferprinted}.

\begin{figure}[!htb]
    \centering
    \includegraphics[width=\textwidth]{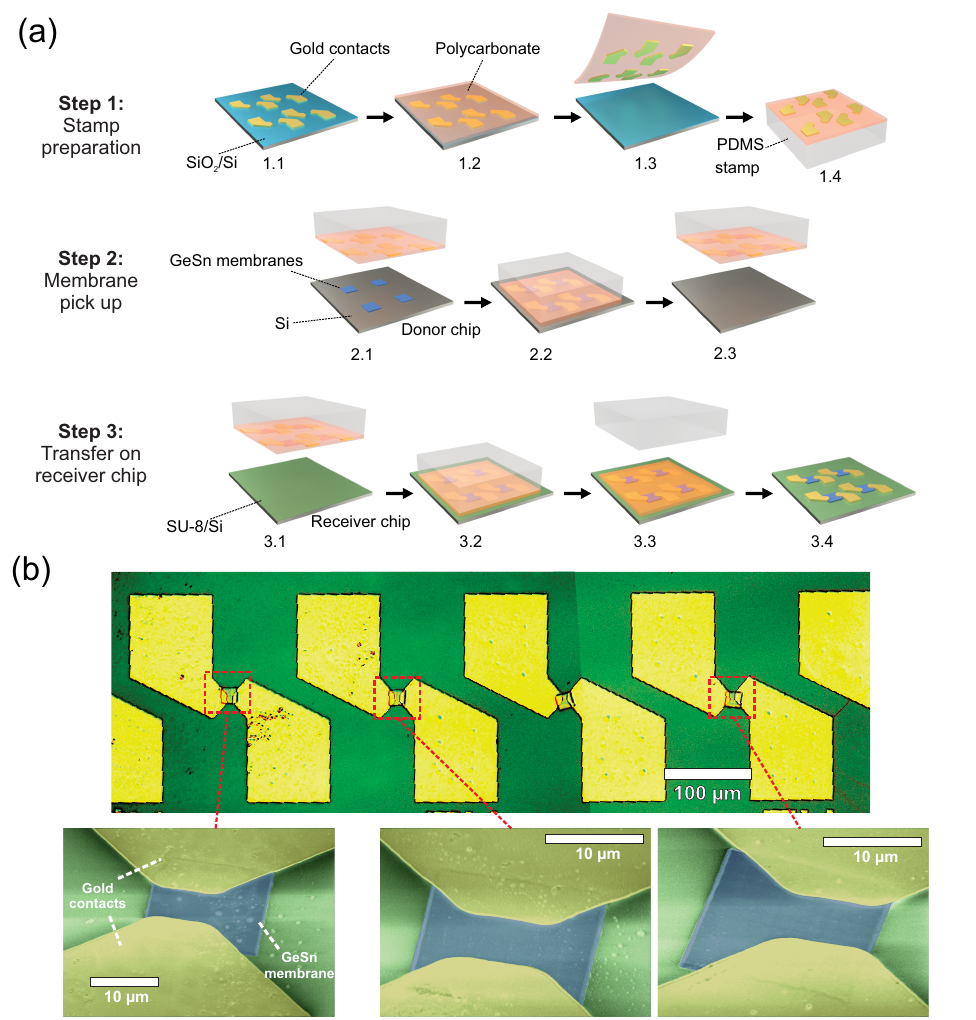}
    \caption{Membrane photodetector fabrication via transfer printing. (a) Transfer printing steps. (b) Optical microscope image of multiple membrane devices (top) along with SEM images of selected devices (bottom).}
    \label{fig:3}
\end{figure}

\noindent {\bf Device characterization}

\begin{figure}[!htb]
    \centering
    \includegraphics[width=\textwidth]{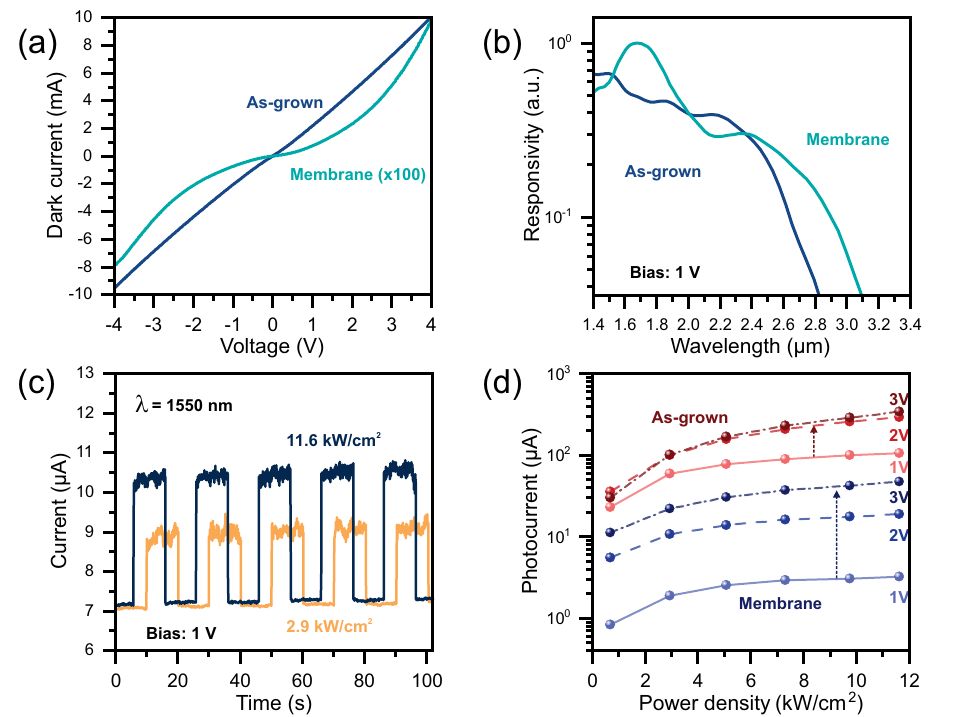}
    \caption{Photoconductive device characterization. (a) Dark current of the contacted membrane and as-grown samples as a function of voltage. (b) Spectral responsivity of the membrane and as-grown devices under a bias of 1 V. (c) Time-dependent photocurrent response of a membrane device under a bias of 1 V and illuminated under different power densities using a $1.55 \,\mu$m laser. (d) Photocurrent as a function of power density and of applied bias.}
    \label{fig:4}
\end{figure}

Electrical characterization of the Ge$_{0.89}$Sn$_{0.11}$ photoconductive (PC) devices was conducted using a Keithley 4200A semiconductor parameter analyzer to assess their performance. The transferred membrane devices are compared to as-grown Ge$_{0.89}$Sn$_{0.11}$ PC devices to highlight the advantages of strain relaxation and transfer onto a dielectric-caped substrate. Dark current as a function of voltage, as shown in Fig. 4(a), is systematically measured from -4 V to 4 V for both the as-grown and the membrane samples. A pronounced reduction by two orders of magnitude in the dark current is observed in the membrane device compared to its as-grown counterpart. The Schottky behavior observed in the dark current curve of the membrane device is most likely due to the passivation resulting from the underetching of Ge using a CF$_4$ plasma, leading to a suppression of the number of surface states through the formation of Ge--F covalent bonds and subsequent alleviation of Fermi level pinning~\cite{wu2011impact}. 

The spectral responsivity for both as-grown and membrane devices is presented in Fig. 4(b). The acquisition of the responsivity spectra was carried out using a Fourier-transform infrared (FTIR) spectrometer (Bruker Vertex 70), with a calibrated tungsten lamp serving as a MWIR source. The MWIR light is focused onto the device using a $40 \times$ microscope reflective objective lens. The photocurrent was measured under a bias of 1 V and using a lock-in technique synchronized to an FTIR step scan to acquire the spectral intensity response of the measured PC devices. The spectral responsivity can then be determined from the measured intensity spectra using a calibrated InSb photodetector. A responsivity of \qty{0.3}{\milli\ampere/\watt} is measured at $1.55 \,\mu$m for the membrane device, and the cutoff wavelength is increased from $2.8 \,\mu$m to $3.1 \,\mu$m relative to the as-grown device. The measured wavelength cutoff values are similar to the absorption cutoff values computed using 8-band $k\! \cdot\! p$ (Fig. 1(d)) at their respective strain values within less than $\sim$ 4\% uncertainty. The latter is strongly dependent on the bowing parameters used in the model~\cite{daligou2023radiative}.

The transient response of the membrane photodetector is investigated under a bias of 1 V using a 1550 nm laser, as shown in Fig. 4(c). A distinctive increase in the total current is observed as the illumination is varied over several cycles at a fixed power density. As the power density increases from \qty{2.9}{\kilo\watt/\centi\square\meter} to \qty{11.6}{\kilo\watt/\centi\square\meter}, the collected photocurrent increases from $1.9 \,\mu$A to $3.2 \,\mu$A. Fig. 4(d) presents the photocurrent measured as a function of power density for both as-grown and membrane samples at biases ranging from \qty{1}{\volt} to \qty{3}{\volt}. A notable increase in the photocurrent is noted with rising incident light power density and applied bias. Specifically, the Ge$_{0.89}$Sn$_{0.11}$ membrane photodetector exhibits an increase in responsivity at \qty{1550}{\nano\meter} from \qty{0.3}{\milli\ampere/\watt} to \qty{8.8}{\milli\ampere/\watt} as the bias is increased from \qty{1}{\volt} to \qty{3}{\volt}. In contrast, the as-grown sample shows an increase from \qty{13.8}{\milli\ampere/\watt} to \qty{89}{\milli\ampere/\watt} over the same bias range.

\begin{figure}[!htb]
    \centering
    \includegraphics[width=\textwidth]{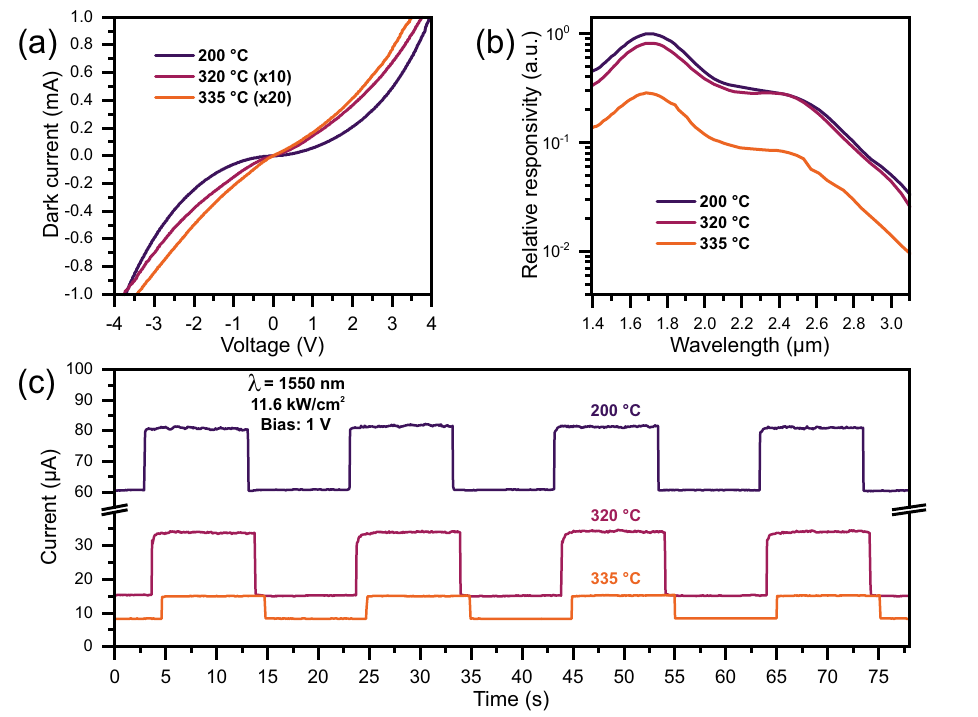}
    \caption{Device characterization after chemical treatment and annealing at different temperatures. (a) Dark current as a function of voltage. (b) Spectral responsivity normalized to the measurement made with the annealing at \qty{200}{\celsius}. (c) Time-dependent photocurrent response under a bias of 1 V and illuminated by a $1.55 \,\mu$m laser at a power density of \qty{11.6}{\kilo\watt/\centi\square\meter}.}
    \label{fig:5}
\end{figure}

The responsivity of the membrane devices can potentially be improved by removing the residual non-volatile species from the release step using cleaning solutions~\cite{gupta2013highly} and by annealing the contacts~\cite{keum2013electrical}. It is expected that Sn should start segregating in Ge$_{0.89}$Sn$_{0.11}$ at an annealing temperature close to the growth temperature~\cite{mukherjee2021atomic}, which is an additional constraint in post-growth processing. In this regard, new devices were fabricated to investigate the influence of surface treatments on the membranes and the annealing of the contacts. To isolate these effects, the process flow is modified to first transfer the membranes to the host chip without transferable contacts. Subsequently, the chip undergoes immersion in a mixture of water and hydrochloric acid (4 H$_2$O : 1 HCl) for 10 s, followed by rinsing in water. The exfoliated contacts are then transferred in a second transfer printing step thereafter. As the polycarbonate film is melted at \qty{200}{\celsius}, membranes and contacts are initially annealed at this temperature for 10 min. The dark current and the responsivity are measured before and after the device annealing at \qty{320}{\celsius} and \qty{335}{\celsius} close to the growth temperature.

The obtained device exhibits a responsivity of 3.1 mA/W at $1.55 \,\mu$m, representing a tenfold increase in comparison to untreated devices. This enhancement may be attributed to the partial removal of residual by-products that could have created a barrier for the carriers. As the annealing temperature increases, the magnitude of the dark current decreases as depicted in Fig. 5(a). The dark current decreases by a factor of 10 after annealing at \qty{320}{\celsius} and by an additional factor of 2 after annealing at \qty{335}{\celsius} while gradually leaning towards an Ohmic behavior. Although the spectral responsivity maintains a consistent wavelength range at different annealing temperatures, a reduction in the responsivity is observed when compared to the initial annealing carried out at \qty{200}{\celsius}, as shown in Fig. 5(b). Specifically, it diminishes to 2.6 mA/W and 0.9 mA/W after annealing at \qty{320}{\celsius} and \qty{335}{\celsius}, respectively.

The time-dependent response of the device under a 1 V bias and illumination by a $1.55 \,\mu$m laser at a power density of \qty{11.6}{\kilo\watt/\centi\square\meter} is presented in Fig. 5(c) for different annealing temperatures. Remarkably, a reduction in both dark current and photocurrent is observed with increasing annealing temperatures. This reduction in dark current is mainly attributed to an enhancement in the crystalline quality of Ge$_{0.89}$Sn$_{0.11}$ through annealing, likely reducing vacancy-like defects capable of trapping charge carriers~\cite{assali2019vacancy}. However, an increase in contact resistivity may also contribute to the reduction with rising annealing temperatures as less photocurrent is being collected. These factors appear to be competing, highlighting the need for defect removal in the microstructure with a minimal impact on contact resistivity.

\section*{Conclusion}

In summary, Ge$_{0.89}$Sn$_{0.11}$ membrane photodetectors were developed and characterized showing room-temperature operation in the mid-infrared range. The control of Sn content and lattice strain throughout the epitaxial growth yielded highly relaxed alloys leading to released membranes without any significant structural bowing, thus facilitating the large-scale transfer of membrane detector arrays. Indeed, the fabrication process based on transfer printing was improved by using an SU-8 adhesive layer and transferable gold contacts. These improvements enabled the transfer of multiple membrane-based photoconductive devices in a single transfer printing step. Furthermore, the effects of annealing and chemical treatment on the photodetectors were explored to identify avenues to improve the device's performance. During the release, the membranes were subjected to an additional relaxation of the residual strain inducing an extension of the cutoff wavelength from $2.8 \,\mu$m to $3.1 \,\mu$m. Moreover, the transferred detectors were found to exhibit two orders of magnitude reduction in the dark current as compared to detectors made of as-grown layers. Our findings not only underscore the potential of Ge$_{1-x}$Sn$_x$ membranes as a robust group IV platform for MWIR photodetection and sensing but also emphasize the ongoing progress in their practical implementation. A platform based on group IV semiconductors benefits from compatibility with silicon, which is vital for the development of large-scale MWIR components.

%%% Textwidth in inches: 6.50128 in
\section*{Acknowledgments}

The authors thank J. Bouchard for the technical support with the CVD system and the Microfabrication Laboratory (LMF) for the support with the microfabrication steps. O.M. acknowledges support from NSERC Canada (Discovery, SPG, and CRD Grants), Canada Research Chairs, Canada Foundation for Innovation, Mitacs, PRIMA Québec, Defence Canada (Innovation for Defence Excellence and Security, IDEaS), the European Union’s Horizon Europe research and innovation program under grant agreement No 101070700 (MIRAQLS), the US Army Research Office Grant No. W911NF-22-1-0277, and the Air Force Office of Scientific Research Grant No. FA9550-23-1-0763. C.L.-L. acknowledges financial support from NSERC (CGS D) and Fonds de recherche du Québec: Nature and Technologies (FRQNT, doctoral scholarship).

%%%%%%%%%%%%%%%%%%%%%%%%%%  bibliography  %%%%%%%%%%%%%%%%%%%%%%%%%%
\bibliography{main.bib}

\end{document}